\documentclass{aa}  

\usepackage{graphicx}
\usepackage[varg]{txfonts}
\usepackage{xcolor}
\usepackage{longtable,lscape}
\usepackage{microtype}
\usepackage{booktabs}

\begin{document}

   \title{Investigating the connection between $\gamma$-ray activity and relativistic jet in 3C\,273 during 2015--2019}

   \titlerunning{$\gamma$-ray emission from the jet of 3C\,273}

   \author{Dae-Won Kim\inst{1},
          Sascha Trippe\inst{1}, and
          Evgeniya V. Kravchenko\inst{2,3}
          }
   \authorrunning{D.-W. Kim et al.}

   \institute{Department of Physics and Astronomy, Seoul National University, Gwanak-gu, Seoul 08826, Korea\\
              \email{dwkim@astro.snu.ac.kr}
         \and
             INAF Istituto di Radioastronomia, Via P. Gobetti, 101, Bologna, 40129, Italy
         \and
             Astro Space Center, Lebedev Physical Institute, Russian Academy of Sciences, Profsouznaya st., 84/32, Moscow, 117997, Russia
             }

   \date{Received Month xx, year; accepted Month xx, year}
 
  \abstract
   {Due to its powerful radiation over the entire electromagnetic spectrum and its radio jet activity, the blazar 3C\,273 offers the opportunity to study the physics of $\gamma$-ray emission from active galactic nuclei. Since a historically strong outburst in 2009, 3C\,273 showed relatively weak emission in the $\gamma$-ray band over multiple years. However, recent \emph{Fermi}-Large Area Telescope observations indicate elevated activity during 2015--2019.
   }
   {We aim at constraining the origin of the $\gamma$-ray outbursts towards 3C\,273 and investigate their connection to the parsec-scale jet.
   }
   {We generate \emph{Fermi}-LAT $\gamma$-ray light curves with multiple binning intervals and study the spectral properties of the $\gamma$-ray emission. Using a 3\,mm ALMA light curve, we study the correlation between radio and $\gamma$-ray emission. Relevant activity in the parsec-scale jet of 3C\,273 is investigated with 7\,mm VLBA observations obtained close in time to notable $\gamma$-ray outbursts.
   }
   {We find two prominent $\gamma$-ray outbursts in 2016 (MJD 57382) and 2017 (MJD 57883) accompanied by mm-wavelength flaring activity. The $\gamma$-ray photon index time series show a weak hump-like feature around the $\gamma$-ray outbursts. The monthly $\gamma$-ray flux--index plot indicates a transition from softer-when-brighter to harder-when-brighter at $1.03\times10^{-7}$\,ph\,cm$^{-2}$\,s$^{-1}$. A significant correlation between the $\gamma$-ray and mm-wavelength emission is found, with the radio lagging the $\gamma$-rays by about 105--112 days. The 43\,GHz jet images reveal the known stationary features (i.e., the core, $S1$, and $S2$) in a region upstream of the jet. We find indication for a propagating disturbance and a polarized knot between the stationary components around the times of both $\gamma$-ray outbursts.
   }
   {Our results support a parsec-scale origin for the observed $\gamma$-ray elevated activity, suggesting association with standing shocks in the jet.
   }


   \maketitle
%

\section{Introduction}
\label{sec:s1}

Blazars, a subclass of active galactic nuclei (AGN), are arguably the most energetic persistent objects in the Universe. They are characterized by powerful non-thermal emission through the entire electromagnetic spectrum, rapid variability and strong polarization \citep{trippe2012}. This phenomenology can be explained by the presence of a relativistic jet whose emission is subject to Doppler boosting due to a small viewing angle between the jet axis and the line of sight (e.g., \citealt{bland2019}). The relativistic jets are responsible for the formation of the observed spectral energy distributions (SEDs) of blazars \citep{chen2018, lewis2018, meyer2019}. The standard model of blazar SEDs (i.e., leptonic scenarios) predicts that the jets radiate via two main processes: synchrotron radiation at radio to UV/X-ray and inverse Compton scattering (IC) at X-rays to $\gamma$-rays (e.g., \citealt{potter2013, piano2018, liodakis2019}; but see also \citealt{hess2019}, for discussion of its limitations and alternative models). Due to inadequate spatial resolution of high energy telescopes, however, our understanding of the high-energy emission is limited, and the site of its production is a matter of active debate.

$\gamma$-ray emission from blazars is known to vary on short time-scales ranging from minutes to days (e.g., \citealt{nalewajko2013, petropoulou2015, meyer2019}) and occasionally shows distinct spectral variations such as hardening or softening (e.g., \citealt{rani2013b, kim2018d, shah2019}; but see also \citealt{nalewajko2013, paliya2015}, for limitations in the spectral analysis). Furthermore, multi-waveband studies reported strong positive correlations between $\gamma$-ray emission and emission at lower frequencies (e.g., \citealt{algaba2018, prince2019}). The observations provide hints at the emission physics: a small or narrow emission region, variations in acceleration/cooling processes, and spatial separation between emitting regions at different observing frequencies. The $\gamma$-ray/radio correlations (e.g., \citealt{pushkarev2010, leon2012, ramakrishnan2015, lico2017}) particularly played an important role to link the $\gamma$-ray production site to the VLBI radio core which is generally identified as the brightest, compact, synchrotron self-absorbed feature in the Very Large Baseline Interferometry (VLBI) images of blazar jets (\citealt{kovalev2009, kim2018j}; but see also \citealt{lee2016b}, for optically thin spectra of blazars at mm-wavelengths in dominance of the core). In addition, the absence of $\gamma$-ray absorption by broad-line region (BLR) photons further supports a location of the $\gamma$-ray dissipation zone downstream the relativistic jet (i.e., $>$~$10^{4}~R_{s}$, with $R_{s}$ being the Schwarzschild radius), where the parsec scale radio core appears (e.g., \citealt{harris2012, costamante2018, meyer2019}; see also \citealt{jorstad2013, kravchenko2016, kim2018d}, for VLBI studies consistent with the idea). However, the BLR region, which is closer to the central black hole, is also a well-known $\gamma$-ray production site as revealed by observations of several objects (e.g., \citealt{rani2013b, berton2018}; but see also \citealt{hodgson2018, rani2018}, for discussion of multiple $\gamma$-ray sites).

The flat-spectrum radio quasar 3C\,273 is one of the most extreme blazars, showing strong and flaring radiation throughout the electromagnetic spectrum. 3C\,273 displays a bright, extended relativistic jet in cm/mm VLBI images, and is known to be a powerful high energy emitter \citep{bruni2017}, making it one of the best sources to study the nature of the $\gamma$-ray emission from blazars. A number of recent studies \citep{rani2013b, chidiac2016, lisakov2017} indicate a region close to the jet apex, rather than the 7\,mm core, to be the place of origin of bright $\gamma$-ray outbursts. Utilizing the energy dependence of electron cooling times, one can determine the site of the IC scattering which generates the $\gamma$-ray emission \citep{dotson2012}. This provides information on the place of origin of the IC seed photons -- BLR or dusty torus -- and thus provides a distance scale -- subparsec or parsec -- for the $\gamma$-ray production in the jet \citep{dotson2015, coogan2016}. A common assumption in multiple scenarios envoked to explain blazar $\gamma$-ray flaring activity is the presence of a plasma blob moving downstream the jet; these structures traveling along the jet are observed frequently. As it propagates along the jet, the blob can pass, and interact with standing shocks (e.g., \citealt{wehrle2016, hodgson2017}; see also \citealt{bottcher2019}, for discussion of $\gamma$-ray flares with different origins). The latter may appear as stationary features (e.g., \citealt{gomez1997, hervet2016}) in the VLBI jets.

In this study, we investigate the 2015--2019 $\gamma$-ray emission of 3C\,273 associated with its relativistic jet to explore the origin of $\gamma$-ray outbursts. The observations and data are described in Section~2. In Section~3, we present our results and analysis. Finally, we discuss and conclude the observed phenomena in Section~4 and Section~5, respectively. We use the following cosmological parameters: $H_{0}$ = 71\,km\,Mpc\,s$^{-1}$, $\Omega_{\Lambda}$ = 0.73, and $\Omega_{m}$ = 0.27, corresponding to an angular scale of 2.71\,pc/mas at the redshift of 3C\,273, $z=0.158$ \citep{strauss1992}.

\section{Observations}
\subsection{Fermi-LAT}
\label{sec:s2.1}

We analyze Pass 8 $\gamma$-ray data obtained by the \emph{Fermi}-LAT \citep{atwood2009}. The data are calibrated following the standard unbinned likelihood procedure\footnote{\url{https://fermi.gsfc.nasa.gov/ssc/data/analysis/documentation/}}. We select SOURCE class events at 0.1--300\,GeV measured between 2015 January 1 and 2018 December 10 (MJD: 57023--58462). Filter parameters \texttt{DATA\_QUAL>0 \&\& LAT\_CONFIG==1} and \texttt{zmax=90$^{\circ}$} were selected for the good time intervals and minimize the Earth limb $\gamma$-ray contamination, respectively. We defined a region of interest (ROI) of $15^{\circ} \times 15^{\circ}$ centered at 3C\,273, and include all sources in the 4FGL catalog (i.e., \citealt{fermi2019}) within the ROI. To take into account diffuse background sources, the Galactic diffuse emission \texttt{gll\_iem\_v07} and isotropic background emission \texttt{iso\_P8R3\_SOURCE\_V2\_v1} templates are applied. The significance of $\gamma$-ray signals is evaluated with the maximum likelihood test statistic (TS). At first, we optimize the background model using the ScienceTools (\texttt{v11r5p3}). We perform a maximum likelihood fit to the data covering half of the whole period leaving the spectral parameters of all sources free. For the two diffuse backgrounds, the normalizations, including the index of the Galactic diffuse emission, are left free. The sources with TS $<$ 10 in the background model obtained from the first optimization are removed, then a second optimization is performd with the updated model. To produce the final $\gamma$-ray light curves, we leave the spectral parameters free for sources within 3$^{\circ}$ from the ROI center, plus the normalizations for the diffuse backgrounds. All the other parameters are fixed to the results of the second optimization. The spectral model of 3C\,273 is assumed to be a power-law model\footnote{\url{https://fermi.gsfc.nasa.gov/ssc/data/analysis/scitools/source_models.html}} defined as $dN/dE \propto E^{+\Gamma}$ with $N$ being the number of photons, $E$ being the photon energy, and $\Gamma$ being the photon index. We compute $2\sigma$ upper limits for $\gamma$-ray signals detected with TS $<$ 9 or $\Delta F_{\gamma}/F_{\gamma} > 0.5$, where $F_{\gamma}$ and $\Delta F_{\gamma}$ indicate the flux and its error, respectively. As binning intervals for the $\gamma$-ray light curves, we select 30 and 7\,days for the full time range (i.e., 2015--2019), and one day for flaring periods to provide a ``zoom in'' view. Given the average flare duration of 12\,days reported in \citet[][]{abdo2010c}, weekly and monthly time bins are appropriate to describe the global $\gamma$-ray activity of blazars, as also noted in previous studies \citep[][]{rani2013b, chidiac2016, meyer2019}. 3C\,273 was not very bright during 2015--2019. Thus, we prefer an interval of at least 7\,days to achieve a high signal-to-noise ratio. Weekly binning also coincides with the average sampling interval of the ALMA data used in this study (see Section~\ref{sec:s3.3}).

\subsection{ALMA Band 3}
\label{sec:s2.2}

We make use of a radio light curve of 3C\,273 provided by the Atacama Large Millimeter/submillimeter Array (ALMA)\footnote{\url{https://almascience.eso.org/alma-data/calibrator-catalogue}} spanning from 2015 January to 2018 December. The flux data were obtained at ALMA band 3 (84--116\,GHz), with the majority of them have been taken at 91 and 103\,GHz. In the case of multiple flux measurements on a single day, the data point with the smallest error is considered here. Further details of the observations can be found in \citet{bonato2018}.

\begin{figure*}  
\centering
 \includegraphics[width=15cm]{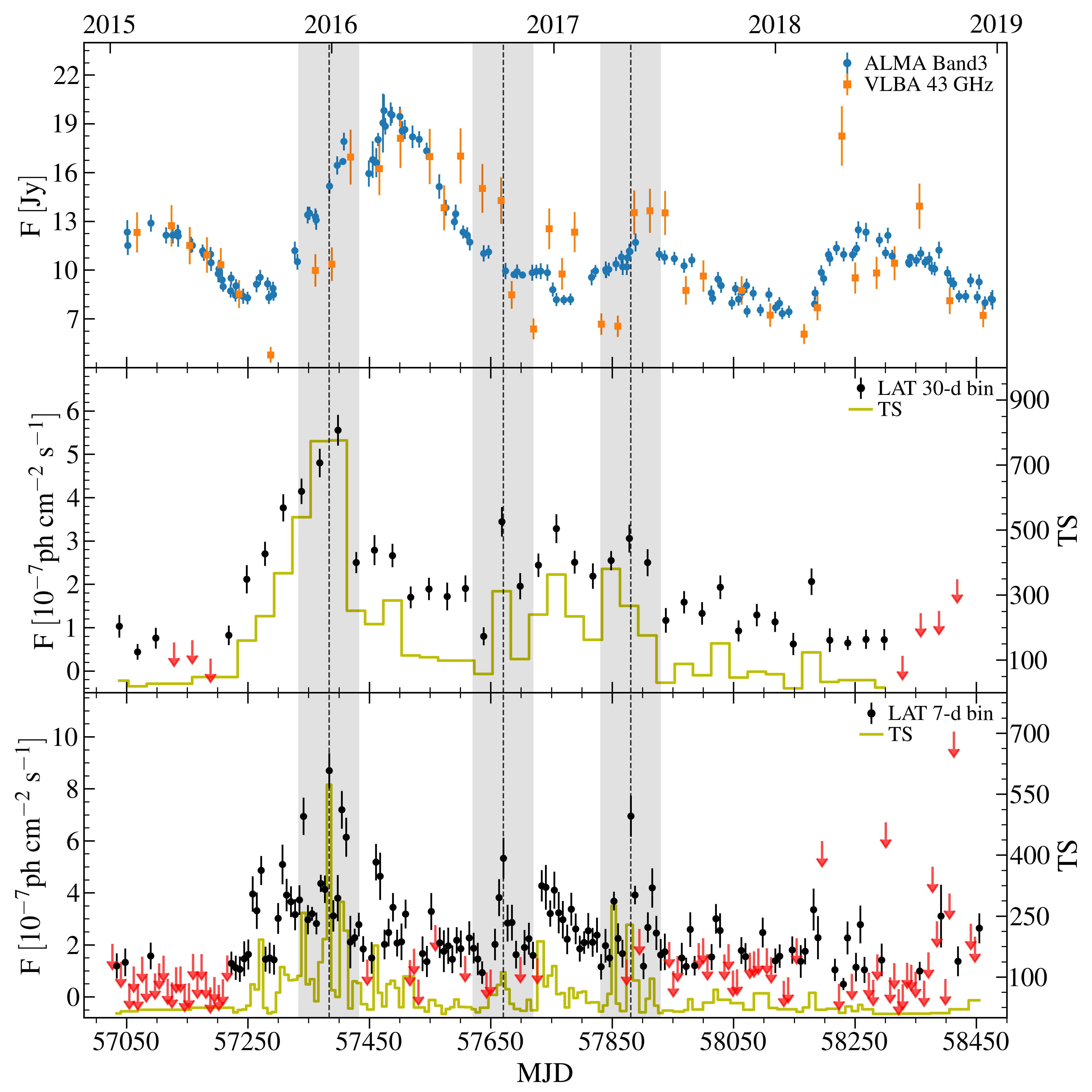}
 \caption{$\gamma$-ray and radio light curves of 3C\,273 in 2015--2019. From top to bottom: ALMA band 3 ($\sim$100\,GHz) overlaid on VLBA 43\,GHz total flux, \emph{Fermi}-LAT fluxes using monthly binning, and \emph{Fermi}-LAT fluxes using weekly binning. For the $\gamma$-ray light curves, 2\,$\sigma$ upper limits are indicated by red downward arrows. The vertical dashed lines show three $\gamma$-ray outbursts identified in the weekly light curve. Each of the shaded areas spans 100 days centered at the peak of a $\gamma$-ray outburst.
}
 \label{fig:f1}
\end{figure*}

\subsection{VLBA 43 GHz}
\label{sec:s2.3}

The VLBA-BU-BLAZAR program (i.e., \citealt{jorstad2016}) monitors bright $\gamma$-ray blazars (34 blazars and 3 radio galaxies) monthly with the Very Long Baseline Array (VLBA) at 43\,GHz. We used fully calibrated data for 3C\,273, available publicly\footnote{\url{https://www.bu.edu/blazars/VLBAproject.html}}. Data are imaged with the software package \texttt{Difmap} \citep{shepherd1997}. The image analysis is performed with several datasets observed close to the time of two $\gamma$-ray outbursts: 2015 December to 2016 April and 2017 April to 2017 August (10 epochs in total). We consider a conservative accuracy of flux densities of 10\%. For the positions of the brightest and compact knots, $\sim$1/10 of the synthesized beam dimensions are used (e.g., \citealt{lister2009}). A full description of the VLBA-BU-BLAZAR data can be found in \citet{jorstad2017}. We further produce linear polarization maps similar to \citet{kim2018d} for all epochs. We also use the CLEANed model components provided by the BU group in their website\footnote{\url{https://www.bu.edu/blazars/VLBA_GLAST/3c273.html}} to investigate the 43\,GHz fluxes during 2015--2019. We follow \citet{lee2016a} to calculate brightness temperatures and resolution limits for observed jet components. The rms noise of a jet component is estimated from an area spanning 3~$\times$~3 beam sizes centered at the position of the component.

\section{Results}
\subsection{Light curves \label{sec:s3.1}}

Figure~\ref{fig:f1} shows the $\gamma$-ray and radio light curves of 3C\,273 from 2015 January 11 to 2018 December 24 (MJD: 57033--58476). The 3\,mm ALMA light curve comprises one major flare and three minor flares during this time, with an average radio flux of 11$\pm$3\,Jy and a minimum of about $\sim$7\,Jy. The major flare spans from mid-2015 to late-2016 with a peak reaching $\sim$20\,Jy on 2016 March 27 (MJD 57474), implying an increase in the flux density by a factor of about 2.5. We also notice the presence of a sub-structure in this flare: an extra peak of $\sim$18 Jy on 2016 January 21 (MJD 57408). Interestingly, it seems that each ALMA flare has sub-structure (``sub-flares''). The other, minor, flares last less than a year with relatively weak peaks below 13\,Jy. The 43\,GHz VLBA fluxes essentially follow the ALMA light curve, though the relatively large cadence ($\sim$30 days) prevents a more exact comparison. We notice that the 43\,GHz fluxes are both higher and lower than the ALMA fluxes at different times. This implies a variable radio spectrum at mm-wavelengths. It is worth noting that a significant fraction of the total VLBI flux, about 1--6 Jy depending on the time, is contributed by the extended jet structure beyond 0.3--0.4\,mas from the core (see Figure~\ref{fig:f7}).

\begin{figure*}
\centering
 \includegraphics[width=\textwidth]{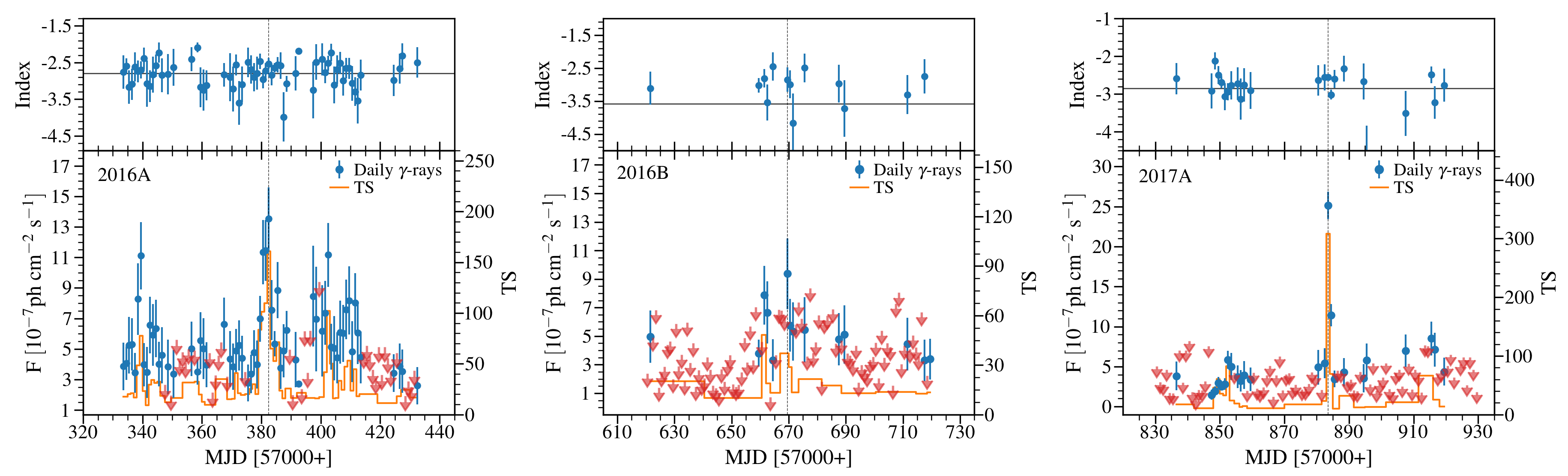}
 \caption{``Zoom in'' $\gamma$-ray light curves of 3C\,273, binned into one-day time bins, for the three $\gamma$-ray flaring periods indicated by the shaded areas in Figure~\ref{fig:f1}. The upper panels in each diagram show the photon indices at the same time resolution. Average photon index values are marked by horizontal solid lines. The vertical dashed lines mark the times of the peaks of the $\gamma$-ray outbursts in the daily light curves.
}
 \label{fig:f2}
\end{figure*}

The monthly and weekly binned $\gamma$-ray light curves yield average fluxes of $(2.0\pm1.2)\times10^{-7}$\,ph\,cm$^{-2}$\,s$^{-1}$ and $(2.6\pm1.4)\times10^{-7}$\,ph\,cm$^{-2}$\,s$^{-1}$, respectively. It is worth noting that 3C\,273 has been in a relatively low-$\gamma$-ray flux state since the historical powerful outburst around 2009 September \citep[see][]{lisakov2017, meyer2019}. However, we find three notable and distinguishable $\gamma$-ray events in our light curves. The first one (2016A) can be found in the monthly light curve with a peak reaching $5.6\times10^{-7}$\,ph\,cm$^{-2}$\,s$^{-1}$ on 2016 January 11 (MJD 57398). This $\gamma$-ray outburst overlaps in time with the major ALMA flare. The other events can be identified more clearly in the weekly light curve. We find the second one (2016B) on 2016 October 9 (MJD 57670) with a peak of $5.3\times10^{-7}$\,ph\,cm$^{-2}$\,s$^{-1}$. The third one (2017A) appeared on 2017 May 7 (MJD 57880) with a peak of $7.0\times10^{-7}$\,ph\,cm$^{-2}$\,s$^{-1}$. These outbursts seems to have 3-mm radio counterparts that are weak compared to the case of the first $\gamma$-ray event. Particularly, the third event is evident only in the light curve with weekly binning. Similarly, the peak of the first $\gamma$-ray outburst can be localized more precisely in the weekly light curve, with a flux of $8.7\times10^{-7}$\,ph\,cm$^{-2}$\,s$^{-1}$ on 2015 December 27 (MJD 57383).

To analyze the $\gamma$-ray flares in more detail and to minimize the impact of binning, we produce $\gamma$-ray light curves with one-day time binning for each outburst \citep[e.g.,][]{marscher2010, wehrle2016}. Figure~\ref{fig:f2} shows the daily $\gamma$-ray light curves which cover the time ranges indicated by the shaded areas in Figure~\ref{fig:f1}. It turned out that the first outburst peaks on 2015 December 26 (MJD 57382) with a flux of $1.4\times10^{-6}$\,ph\,cm$^{-2}$\,s$^{-1}$. For the second and third outburst, the peaks occur on 2016 October 8 (MJD 57669) with a flux of $9.4\times10^{-7}$\,ph\,cm$^{-2}$\,s$^{-1}$ and on 2017 May 10 (MJD 57883) with a flux of $2.5\times10^{-6}$\,ph\,cm$^{-2}$\,s$^{-1}$, respectively. The 2016A event is by far the most prominent one, showing variable enhanced activity around its peak, whereas the $\gamma$-ray activity around the peaks of the 2016B and 2017A events appears to be substantially weaker. We do not see any notable temporal variation in the photon indices derived from the three daily light curves; the average photon index values are $-$2.80, $-$3.58, and $-$2.85, respectively. One can easily identify the first and third outbursts in their daily light curves due to their brightness and high statistical significance, whereas the 2016B outburst is weak and shows relatively low TS. Furthermore, the quality of the BU data obtained on 2016 October 6 is rather poor because two VLBA antennas were not available. Hence, we focus on the 2016A and 2017A $\gamma$-ray outbursts in our further analysis.

\begin{figure}
 \includegraphics[width=\columnwidth]{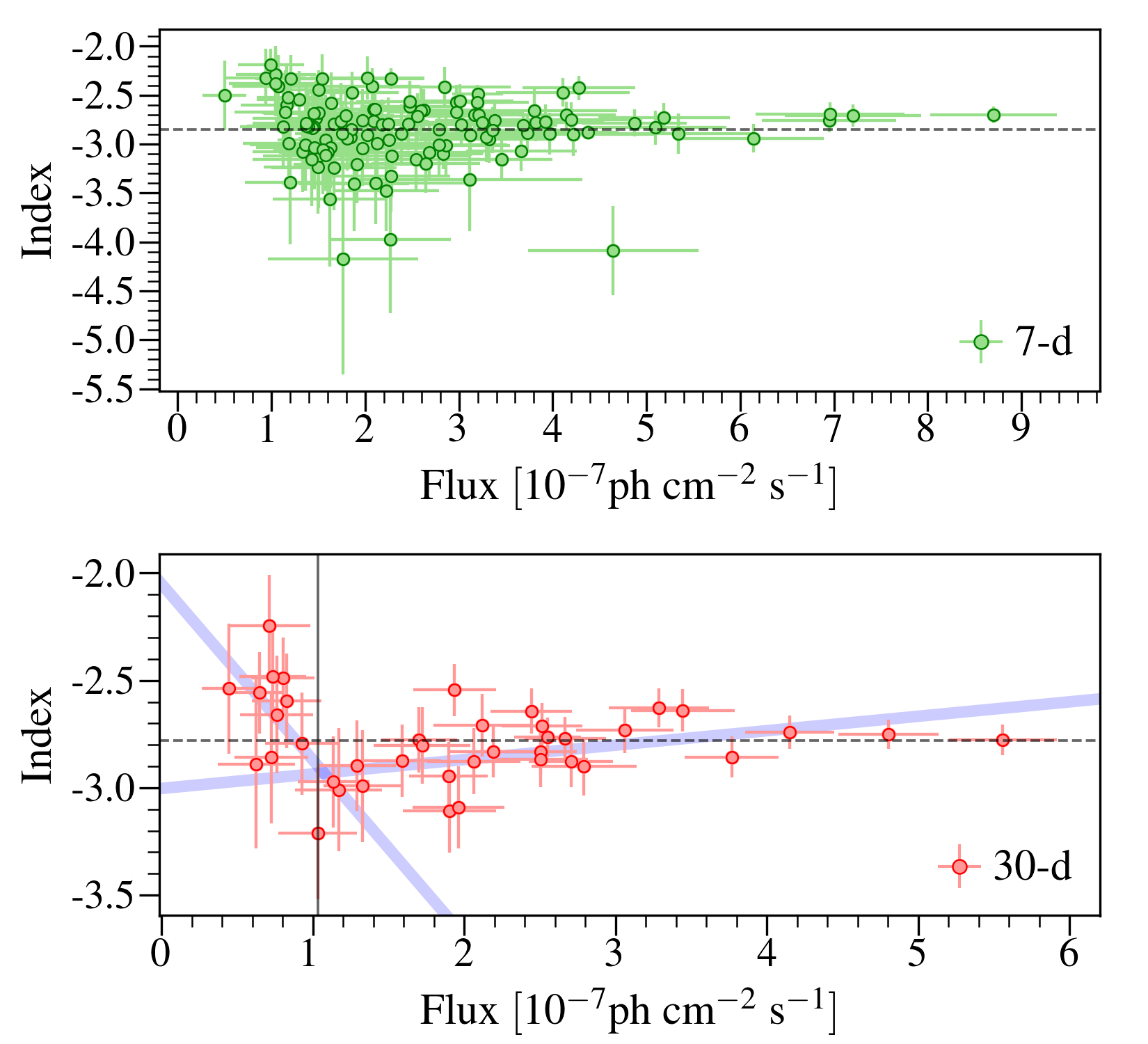}
 \caption{Photon index vs. flux for the weekly binned (\textit{top}) and monthly binned (\textit{bottom}) $\gamma$-ray light curves. The dashed lines indicate the average photon index values. The vertical solid line in the bottom panel marks a break flux value of $1.03\times10^{-7}$\,ph\,cm$^{-2}$\,s$^{-1}$. The light blue lines show the linear regression lines for photon indices above and below the break flux.
}
 \label{fig:f3}
\end{figure}

\subsection{Photon indices from weekly and monthly $\gamma$-ray light curves}
\label{sec:s3.2}

In Figure~\ref{fig:f3}, we present the photon indices derived from the LAT light curves (see Section~\ref{sec:s2.1} for details) as a function of $\gamma$-ray flux. The indices range from $-$4.5 to $-$2.0 for weekly binning, and from $-$3.5 to $-$2.0 for monthly binning. Their averages are $-$2.85 and $-$2.78, respectively. We do not see any correlations (i.e., softer-when-brighter or harder-when-brighter trends) for the weekly $\gamma$-ray fluxes. However, simple visual inspection of the monthly data suggests different trends above and below a flux of about  $1.03\times10^{-7}$\,ph\,cm$^{-2}$\,s$^{-1}$. For fluxes below the threshold, fulx and photon index show a negative correlation with a Pearson coefficient of $-$0.49 at a confidence level of 87\%; above the threshold, there is a positive correlation with a Pearson coefficient of 0.48 at a confidence level of 99\%. Accordingly, the source appears to have been in a harder-when-brighter spectral state at $\gamma$-ray fluxes above $1.03\times10^{-7}$ ph cm$^{-2}$ s$^{-1}$. There is also indication of a softer-when-brighter trend at lower fluxes, though at insufficient confidence level.

\begin{figure}
 \includegraphics[width=\columnwidth]{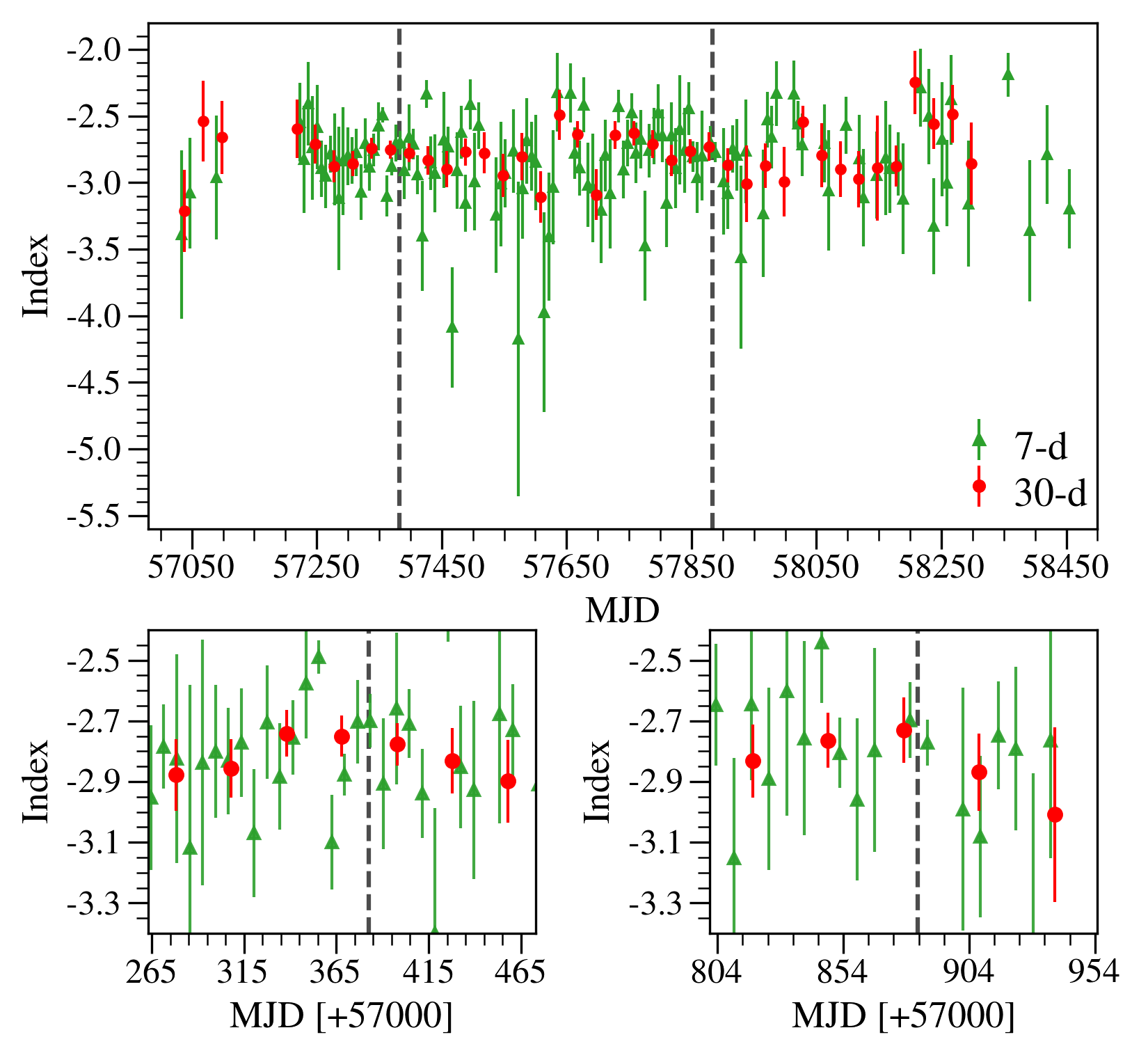}
 \caption{Photon index vs. time of the weekly and monthly $\gamma$-rays. The \emph{top} panel shows the entire time range of our observations. The \emph{bottom} panels provide zoom-in views on the data around the times of the 2016A and 2017A $\gamma$-ray outbursts defined in Figure~\ref{fig:f2} (marked by vertical dashed lines). Errorbars correspond to 1\,$\sigma$ uncertainties.
}
 \label{fig:f4}
\end{figure}

The evolution of the photon index with time during 2015--2019 is shown in Figure~\ref{fig:f4}, using weekly and monthly binning. Overall, the uncertainties of the photon indices are too large for a meaningful quantitative analysis. We therefore limit ourselves to a qualitative discussion of potential patterns or trends in the data. There is indication for gradual temporary increases of the monthly photon indices during the two $\gamma$-ray outbursts. The first hump spans $\sim$180 days (MJD: 57278--57458) with a peak value of $-$2.74 on MJD 57338, while the second hump spans $\sim$120 days (MJD: 57818--57938) with a peak value of $-$2.73 on MJD 57878. Interestingly, the 2016 $\gamma$-ray outburst lags about 44\,days behind the local peak of the first hump, whereas the 2017 outburst coincides with the local peak of the second hump. This might indicate the presence of multiple $\gamma$-ray emitting regions in the jet, thus suggesting the presence of a first dissipation zone which is responsible for these local peaks. 

The weekly indices seem to be fluctuating randomly without showing any noteworthy trends throughout our observations. We attribute this to the combination of typically modest variations in the spectral index \citep[e.g.,][]{abdo2010a, lisakov2017} and low-photon-number statistics which causes a substantial scattering of the photon index values \citep[][]{nandikotkur2007, rani2013b}. The photon statistics improves with stronger source activity \citep[e.g.,][]{paliya2015}. However, 3C\,273 was relatively weak during 2015--2019 (cf. Section~\ref{sec:s3.1}). \citet[][]{nandikotkur2007} found that binning data into longer time intervals can reduce the large errorbars of the spectral index values. Thus, the monthly indices arguably describes the $\gamma$-ray spectral variations in our data better than the weekly ones.

\subsection{Correlation between the radio and $\gamma$-ray light curves}
\label{sec:s3.3}

We employ the discrete correlation function (DCF; \citealt{edelson1988}) to investigate the relationship between $\gamma$-ray and radio emission. The average sampling interval for the ALMA light curve is $\sim$9.8 days. Disregarding the upper limits, the average sampling intervals of the weekly and monthly $\gamma$-ray light curves are $\sim$11.7 and $\sim$32.3 days, respectively. We therefore use time steps of 12 and 33\,days for the DCF analysis for the weekly and monthly binned $\gamma$-ray fluxes, respectively. To determine height and location of a DCF peak, we fit a Gaussian function defined as $DCF(t) = a\,\times$\,exp$\left[-(t-c)^{2}/2w^{2}\right]$ to the DCF durve, with $a$ being the amplitude, $c$ being the time-lag, and $w$ being the width of the Gaussian profile. The statistical significance of the correlation is calculated following \citet{rani2013a}, by calculating the Pearson correlation coefficient for the ALMA data and LAT data and its uncertainty after applying the time shift derived from the Gaussian fit to the DCF curve.

\begin{figure}
 \includegraphics[width=\columnwidth]{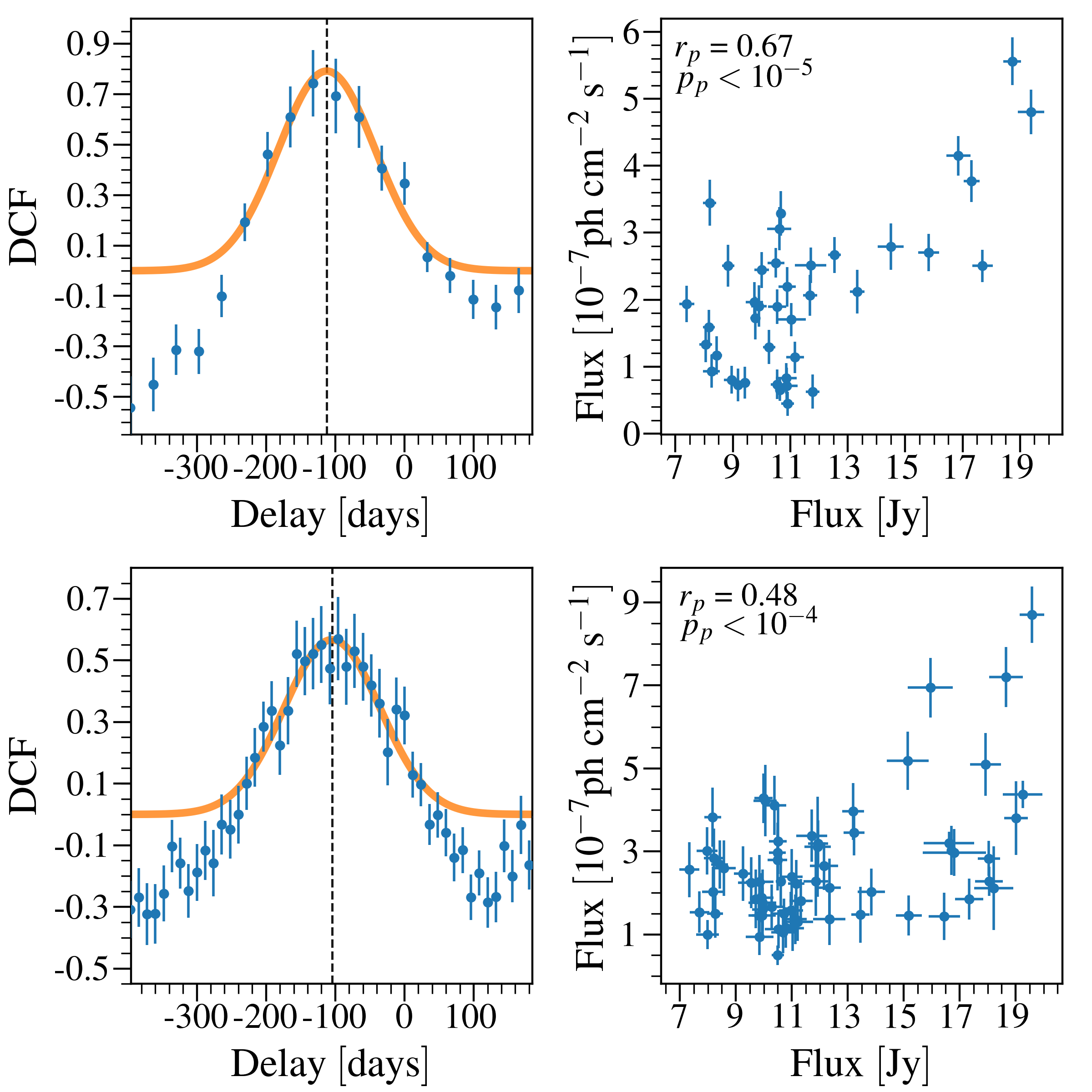}
 \caption{DCF curves and flux--flux plots comparing the 3-mm radio and $\gamma$-ray light curves in 2015--2019, for the monthly $\gamma$-ray fluxes (\textit{top panels}) and the weekly $\gamma$-ray fluxes (\textit{bottom panels}). Orange curves in the left panels indicate the best Gaussian fits to the DCF curves. For the flux-flux plots (right panels), the ALMA light curve has been shifted in time by the delays found from the Gaussian fits. The Pearson correlation coefficients ($r_{p}$) and corresponding \textit{p}-values ($p_{p}$) are shown in the flux-flux plots.
}
 \label{fig:f5}
\end{figure}

\begin{table}
\caption{Results of our correlation analysis for 2015--2019 data.}
\label{tab:tb1}   
\centering         
\begin{tabular}{l r r}   
\toprule
Parameter  &  LAT 7d vs. ALMA  &  LAT 30d vs. ALMA  \\
\midrule
$a^{1}$  &  $0.57\pm0.03$  &  $0.79\pm0.08$  \\
$c$~(days)  &  $-105\pm4$  &  $-112\pm6$  \\
$\mathrm{\left|w\right|}$~(days)  &  $70\pm4$  &  $70\pm6$  \\
$r_{p}$  &  $0.48$  &  $0.67$  \\
$p_{p}$~(\%)  &  $>99.99$  &  $>99.99$  \\
\bottomrule 
\multicolumn{3}{l}{$^1$ Amplitude of DCF curve.}\\
\end{tabular}
\end{table}

The results of the correlation analysis are shown in Figure~\ref{fig:f5} and summarized in Table~\ref{tab:tb1}. The Gaussian fits to the DCF curves find time delays of $-105\pm4$ and $-112\pm6$\,days ($\gamma$-ray leading) for the weekly and monthly binned $\gamma$-ray lightcurves, respectively. These estimates are consistent with each other within the errors. We note that the location of the DCF peak varies considerably (with a standard deviation of $\sim$12.5\,days) when using different time delay bins for the DCF calculation, for both the weekly and monthly $\gamma$-ray fluxes. However, the time delays found from Gaussian fits are only weakly affected by the binning of the DCF (with a standard deviation of $\sim$0.9\,days). Applying the best-fit delays, the Pearson correlation analysis resulted in coefficient values of 0.67 for the pair with the monthly $\gamma$-ray fluxes and 0.48 for the pair with the weekly $\gamma$-ray fluxes, at significance levels of $>$ 99\% for both (see Figure~\ref{fig:f5}).

\begin{figure}
 \includegraphics[width=\columnwidth]{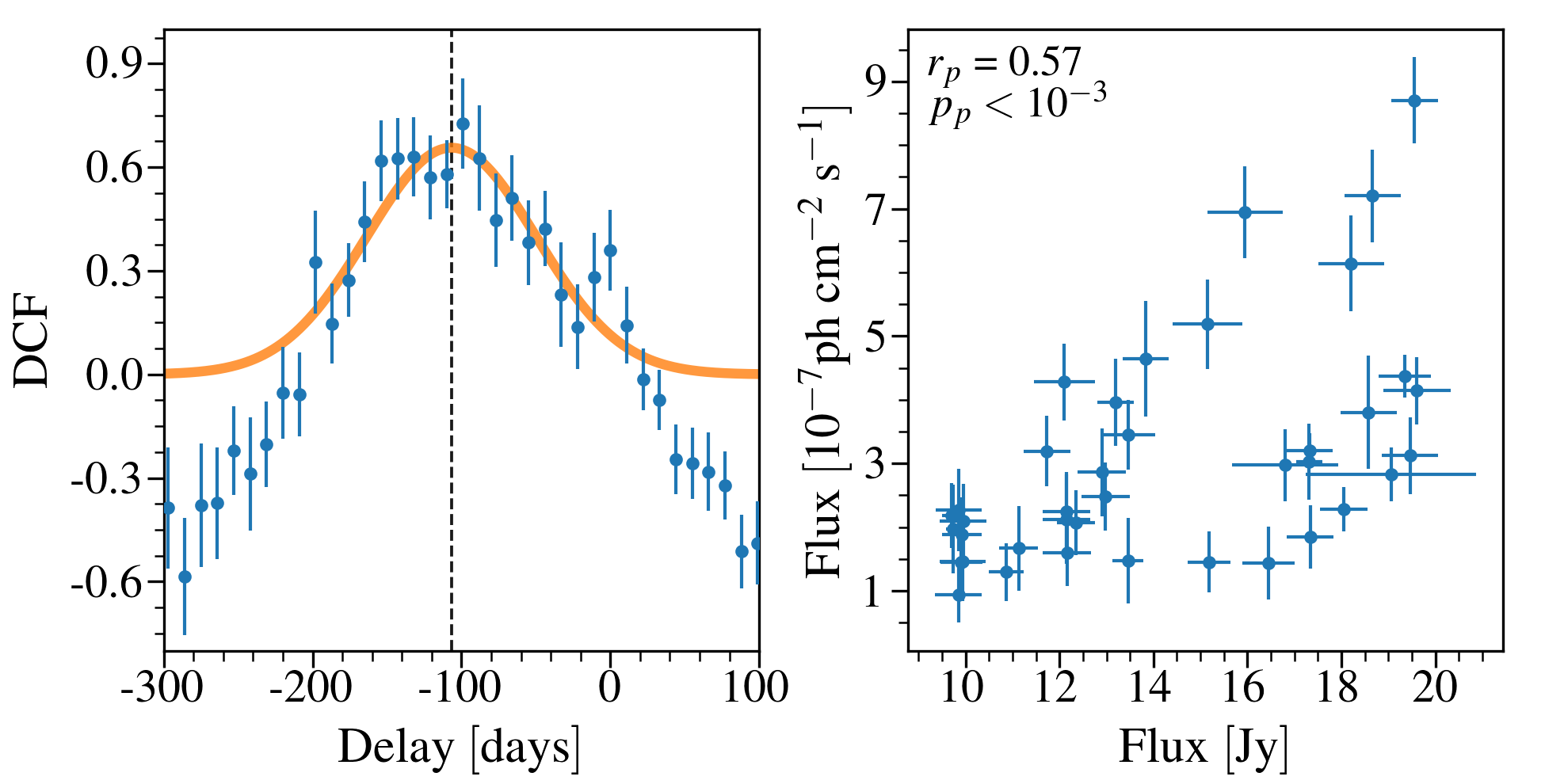}
 \caption{Same as Figure~\ref{fig:f5}, but using only the weekly $\gamma$-ray fluxes in the time range 2015--2017.
}
 \label{fig:f6}
\end{figure}

Simple visual inspection of Figure~\ref{fig:f1} made us suspect that the DCF is dominated by data from the time range 2015--2017. Hence, we repeated our correlation analysis using the parts of the light curves in that time range (see Figure~\ref{fig:f6}). We used the weekly $\gamma$-ray light curve, which includes the fine structure of the flares and is sampled about as densely as the ALMA fluxes. We find a clear correlation comparable to the results in Table~\ref{tab:tb1}, thus confirming the initial impression that the major $\gamma$-ray and radio flares in 2015--2017 dominate the correlation between the full (2015--2019) light curves. Results of the analysis are summarized in Table~\ref{tab:tb2}. We also checked the 2017--2019 data, but found the DCF to be consistent with random fluctuations.

\subsection{Parsec-scale jet near the 43\,GHz core}
\label{sec:s3.4}

\begin{figure}
 \includegraphics[width=\columnwidth]{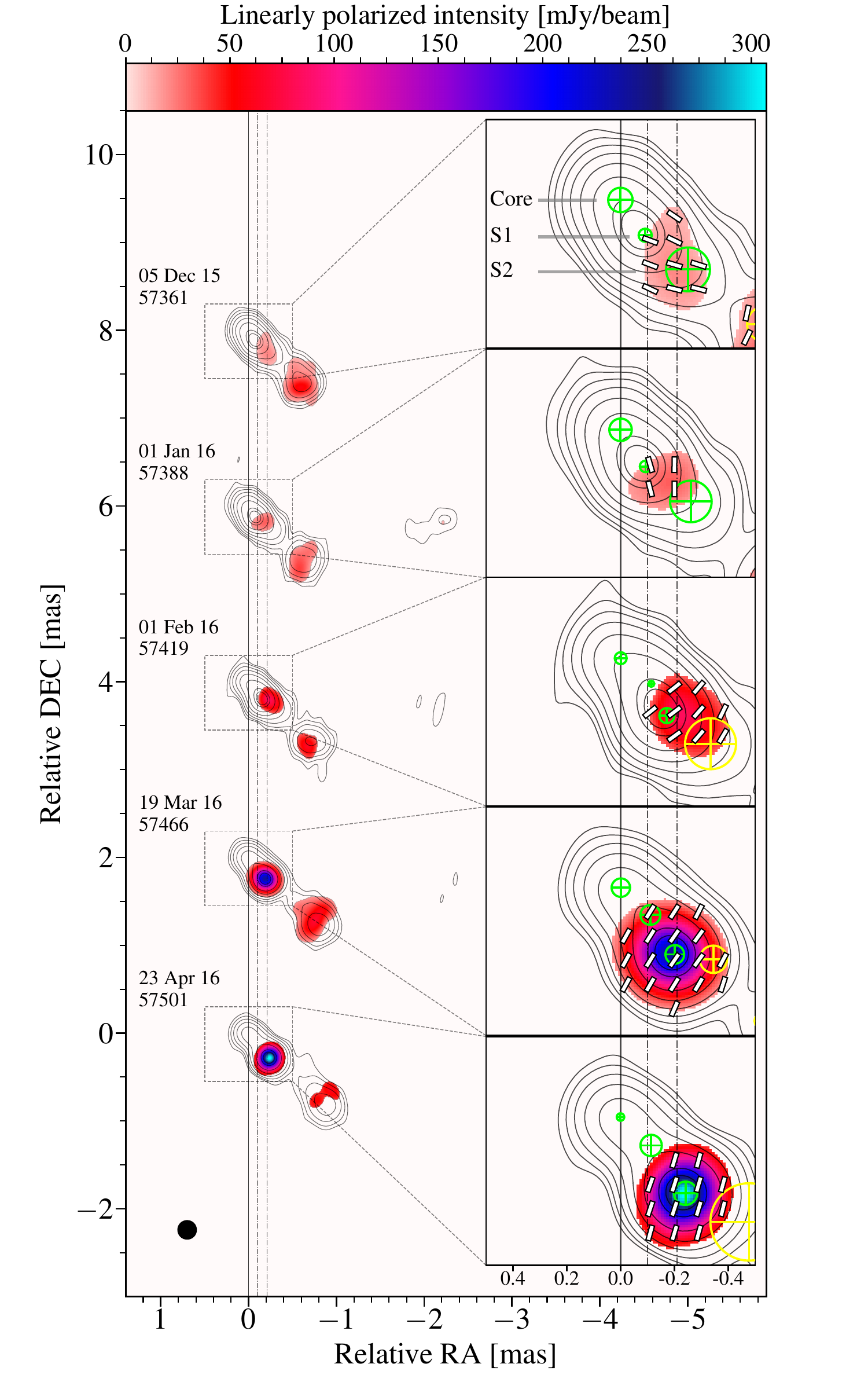}
 \caption{7\,mm VLBA images of 3C\,273 during the 2016 $\gamma$-ray outburst. The contours and color scale represent the total intensity and linearly polarized intensity, respectively. Contour levels are 1\%, 2\%, 4\%, 8\%, ..., 64\%, 80\% of the peak intensity. A zoom-in view on the core region is provided in the subplots. White line segments indicate the electric vector position angle (EVPA) directions uncorrected for Faraday rotation. Circular Gaussian jet components are indicated by green (for the three stationary components) and yellow (for the others) $\oplus$ symbols. The vertical dot-dashed lines show the average positions of the $S1$ (=0.1\,mas) and $S2$ (=0.21\,mas) components projected onto the \textit{x}-axis. The vertical solid line indicates the core position. All maps are restored with a 0.2$\times$0.2\,mas beam (indicated at the bottom left).
}
 \label{fig:f7}
\end{figure}

\begin{figure}
 \includegraphics[width=\columnwidth]{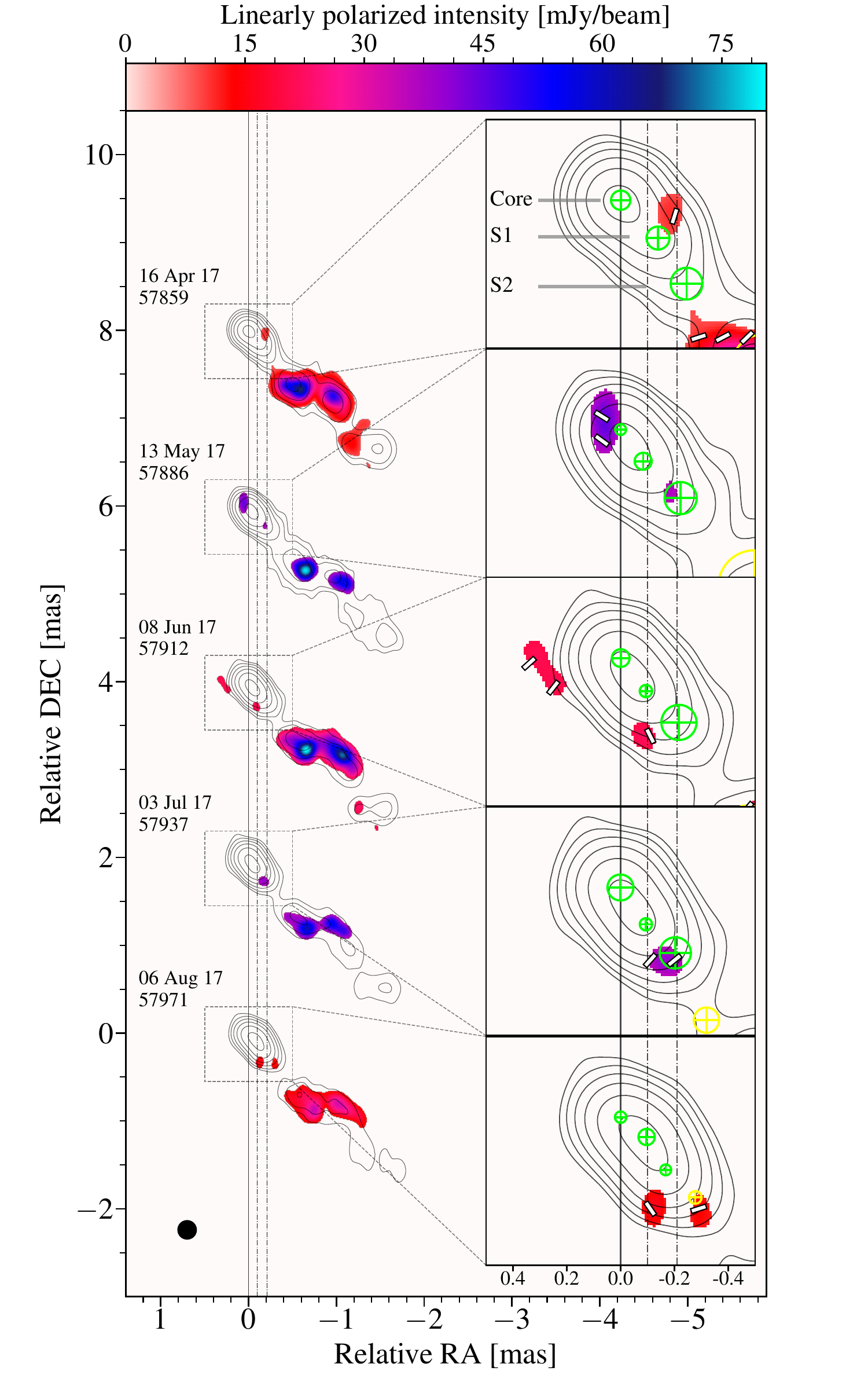}
 \caption{Same as in Figure~\ref{fig:f7}, but for the 2017 $\gamma$-ray outburst. Contour levels are 2.5\%, 5\%, 10\%, ..., 40\%, 80\% of the peak intensity.
}
 \label{fig:f8}
\end{figure}

We investigated the parsec-scale jet of 3C\,273 during times of elevated $\gamma$-ray and radio band activity using VLBA maps. Considering the time-lags between $\gamma$-ray and radio fluxes, we chose two sets of VLBA-BU-BLAZAR observations that span about 4 months after each $\gamma$-ray outburst, shown in Figure~\ref{fig:f7}~and~\ref{fig:f8}. All maps are aligned at the position of the core which is taken to be the upstream end of the jet. In each data set, the structure of the jet was represented by a number of two-dimensional circular Gaussians fitted to the visibility domain. The resultant models consist of 7--10 components, parameters of which are summarized in Appendix~\ref{app:a1}. We found three stationary features: the core and two additional components labelled $S1$ and $S2$ which were already identified in a previous study \citep[i.e.,][]{lisakov2017}, with $S1$ and $S2$ located at about 0.16 and 0.33\,mas from the core, respectively.

\begin{table}
\caption{Results of our correlation analysis for 2015--2017 data.}
\label{tab:tb2}    
\centering          
\begin{tabular}{l c}   
\toprule             
 Parameter &  LAT 7d vs. ALMA  \\
\midrule
$a$  &  $0.66\pm0.05$  \\
$c$~(days)  &  $-107\pm5$  \\
$\mathrm{\left|w\right|}$~(days)  &  $57\pm5$  \\
$r_{p}$  &  $0.57$  \\
$p_{p}$~(\%)  &  $99.98$  \\
\bottomrule
\end{tabular}
\end{table}

In Figure~\ref{fig:f7} and \ref{fig:f8}, a spatial displacement of the 7-mm emission is noticeable during both $\gamma$-ray outbursts. The sequence of maps in Figure~\ref{fig:f7} begins on MJD 57361. The second epoch is just 6 days after the 2016 $\gamma$-ray outburst. In this epoch, the total intensity peak is located at the $S1$ position. Then, the peak moves down to the $S2$ position on MJD 57501 while increasing its intensity up to $\sim$9\,Jy/beam. The proper motion of this displacement is about 0.59\,mas/year which corresponds to an apparent speed of $\sim$6.0\,$c$ (with $c$ being the speed of light). Interestingly, such a displacement of the emission peak can also be seen for the 2017 $\gamma$-ray outburst, though weaker. In Figure~\ref{fig:f8}, the initial position, on MJD 57859, of the peak intensity coincides with the core. Then, until MJD 57886, which is 3 days after the 2017 $\gamma$-ray outburst, it moves downstream to a location between the core and $S1$. The further motion of the  emission peak cannot be clearly recognized until MJD 57937, when the flux increases to $\sim$4.8\,Jy/beam. At this time, both the core and $S1$ have became brighter simultaneously. On MJD 57971, the flux peak is localized at the position of $S1$ and shows a decrease in total intensity. This corresponds to a proper motion of about  0.37\,mas/year, translating into an apparent speed of $\sim$3.8\,$c$. The proper motions are consistent with the radial speeds of the newborn components reported in \citet{lisakov2017}.

\begin{figure}
 \includegraphics[width=\columnwidth]{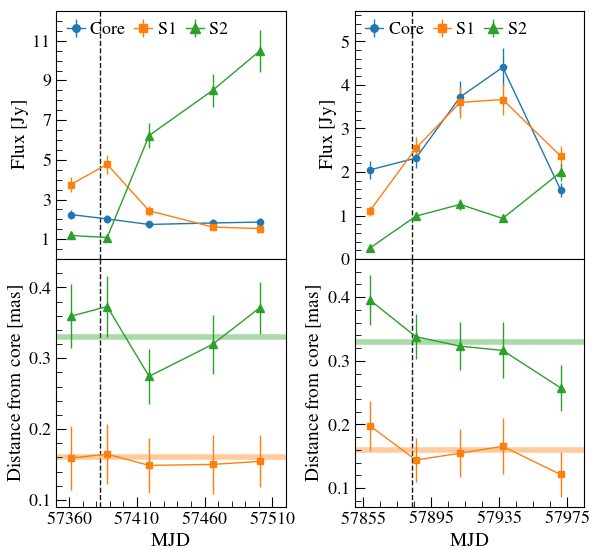}
 \caption{\emph{Left:} Flux density (\textit{top}) and distance from the core (\textit{bottom}) of the core, $S1$, and $S2$ derived from Gaussian modelling of the jet components as shown in Figure~\ref{fig:f7}. \emph{Right:} Same as the left panels, but for the observations in Figure~\ref{fig:f8}. The vertical dashed lines correspond to the times of the 2016 and 2017 $\gamma$-ray outbursts. The average distances of $S1$ and $S2$  (i.e., 0.16 and 0.33\,mas from the core) are indicated by the horizontal solid lines.
}
 \label{fig:f9}
\end{figure}

Figure~\ref{fig:f9} shows the fluxes and distances from the core as function of time for the core, $S1$, and $S2$ as determined from the VLBA data. The $S1$ flux peaks almost simultaneously with the 2016 $\gamma$-ray outburst. The flux of $S2$ increases rapidly, by a factor of 10, just after the 2016 $\gamma$-ray outburst. The core remains quiescent. The 2017 $\gamma$-ray outburst coincides with the rise of a flare of $S1$. In addition, the core fluxes during the first two epochs (i.e., MJD 57859 and 57886) are consistent with each other within the errors, but increase rapidly thereafter. This suggests that the 2017 $\gamma$-ray outburst also coincides with the onset of the core flare. Such a connection can also be made for $\gamma$-ray outburst and the $S2$ flare in 2016. Notably, $S2$ doubles its flux in a month until MJD 57971. As evident in Figure~\ref{fig:f7} and \ref{fig:f8}, the positions of the presumably stationary components show some variation in both epochs studied. $S2$ moved toward the core from MJD 57388 to 57419 by about $\sim$0.1\,mas, accompanied by a rapid increase in its flux. Afterwards, $S2$ slowly moved back to its initial core distance. In the 2017 observations, we notice a tendency for both $S1$ and $S2$ to move toward the core. In general, $S1$ moves less than $S2$.

As the combined total flux density of the core, $S1$, and $S2$ components dominate the mm-wavelength emission of 3C\,273, we can conclude that both the radio flares and the $\gamma$-ray production occur in this region of the jet. We notice that $S2$ is responsible for the major ALMA flare (peaking on MJD 57474). We suggest that the sub-flare (peaking on MJD 57408) within the major ALMA flare could be associated with the activity in $S1$. For the second ALMA flare (around MJD 57888), associated with the 2017 $\gamma$-ray outburst, both the core and $S1$ contribute equally. 

\subsection{Polarization}
\label{sec:s3.5}

Figures~\ref{fig:f7} and \ref{fig:f8} include distributions of linearly polarized intensity overlaid on the total intensity contours. We only mapped out polarized emission that exceeds a conservative significance threshold of 9\,$\sigma$. In the core region, the polarized emission is weak and only observed in a single epoch (MJD 57886), which is consistent with other studies \citep[i.e.,][]{attridge2005, jorstad2005}. In the 2016 observations, we find a polarized knot spatially connected to $S1$ and $S2$ that shows systematic variations in position and intensity. Just after the 2016 $\gamma$-ray outburst (MJD 57388), the polarized knot appears to encounter $S2$ while still covering the $S1$ region. Afterward, it appears to pass through the $S2$ region (MJD 57419) while showing an increase in polarized intensity. This suggests that the emergence of the polarized knot precedes the total intensity peak -- leading us to the conclusion that we observe the downstream propagation of a disturbance in both total and polarized intensity. The polarized intensity peaks in the time from MJD 57466 to 57501 at $\sim$300\,mJy in the $S2$ region.

In 2017, the polarized emission is weaker and shows a more complicated structure. We detect a polarized knot upstream of the core on MJD 57886, right after the 2017 $\gamma$-ray outburst. It disappears before the following observations. Such a behavior might indicate an energetic physical process (e.g., acceleration) happening upstream of the core, thus suggesting an association between the mm-wavelength core and the high energy emission. Weak polarization shows up around $S2$ at two epochs (MJD 57886 and 57937), probably implying physical or geometrical changes in that region.

\section{Discussion}
\subsection{Positional variations of the stationary components}
\label{sec:s4.1}

The 7\,mm VLBA observations reveal the presence of two stationary components in the jet of 3C\,273, $S1$ and $S2$, which were already reported by \citet{lisakov2017}. These features are thought to be multiple recollimation shocks (RCS) that have been predicted by relativistic MHD simulations \citep[e.g.,][]{gomez1995, mizuno2015}. We found a `wiggling' (up- and downstream) motion of $S2$ (see also Figure~2 of \citealt{lisakov2017}, for a similar pattern). $S2$ moved toward the core (inward motion) by about $\sim$0.1 mas from MJD 57388 to 57419, while increasing its flux. It shifted downstream back to its original position afterwards (see Figure~\ref{fig:f9}). Since a new moving knot ($J1$, which appeared on MJD 57419) was passing through the $S2$ region in MJD 57388, both components can appear blended into a single feature, leading to a shift of the centroid downstream of the initial position of $S2$. After the passage, the two components appear to split up, resulting in an apparent inward motion of $S2$. The apparent displacement of a stationary jet component matches the signature expected for a RCS zone breakout (e.g., \citealt{abeysekara2018}). When a moving blob with high kinetic energy starts interacting with the standing RCS localized in the region where the magnetic field becomes unstable, there could be (1) an enhancement of non-thermal emission, (2) a strong instability of the magnetic field configuration (e.g., tearing instability; \citealt{delzanna2016}), and (3) a positional displacement by the underlying flow with increasing kinetic power of the jet \citep{hervet2016}. This scenario also predicts some fast $\gamma$-ray emission induced by magnetic reconnection, though the observed photon indices around MJD 57405 in Figure~\ref{fig:f2} appear too soft \citep{ding2019}. A more conventional interpretation for the $S2$ motions would be an opacity effect \citep[e.g.,][]{lobanov1998}; variations in opacity can occur when flaring components pass through these regions \citep[e.g.,][]{plavin2019}, resulting in an apparent spatial drift.

Yet another possible explanation for the apparent motion of $S2$ in 2016 is the `core shuttle' effect. Changes in the physical state, and thus in the opacity of the core, due to a propagating disturbance may cause a wiggling of the core position. However, the core was in a quiescent state during the time of the $S2$ motion, implying the absence of newly formed jet components. Moreover, a core shuttle should also affect the separation of $S1$ from the core, which is not observed. In 2017, however, a core shuttle might have affected the separations of both $S1$ and $S2$ from the core, which changed simultaneously (see Figure~\ref{fig:f9}) and coincided with an increase in the flux density of the core.

\subsection{On the 2016 $\gamma$-ray outburst}
\label{sec:s4.2}

Our VLBA analysis in Sections~\ref{sec:s3.4} and \ref{sec:s3.5} found that the mm-wavelength activity in 3C\,273 is confined to the most upstream few parsecs of the jet. For the 2016 $\gamma$-ray outburst, the flaring of the stationary components $S1$ and $S2$ strongly suggests that the $\gamma$-ray outburst has a physical connection with these regions. This is further supported by the clear $\gamma$-ray/radio correlation which can be attributed to the major ALMA flare close to the 2016 $\gamma$-ray outburst \citep[see also][for their LCCF result on 3C\,273]{meyer2019}. Moreover, the contemporaneous displacements of the total intensity peak and the polarized knot indicate that a moving disturbance played an important role in the production of the high energy emission.

\begin{figure}
 \includegraphics[width=\columnwidth]{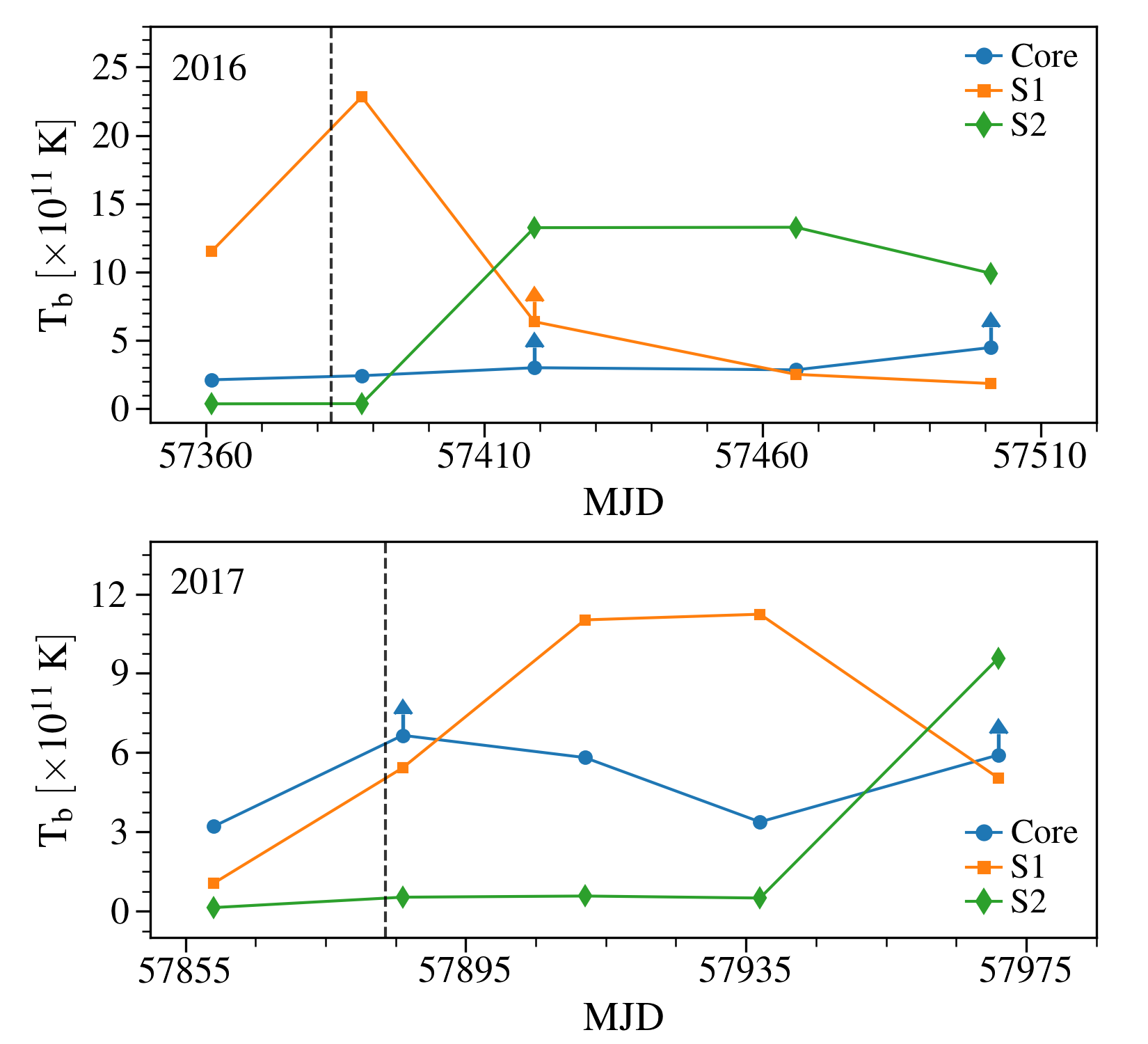}
 \caption{Source-frame brightness temperatures ($T_{b}$) of the core, $S1$, and $S2$. Vertical dashed lines indicate the times of the $\gamma$-ray outbursts. Upward arrows mark limits on $T_{b}$, calculated by using the spatial resolution limit for each component whenever the observed size is smaller than this limit.
}
 \label{fig:f10}
\end{figure}

There has been a considerable number of studies reporting the connection between parsec-scale radio jets and $\gamma$-ray outbursts: $\gamma$-ray outbursts tend to be accompanied by radio flares and strongly polarized jet features \citep{agudo2011a, jorstad2013, wehrle2016, kim2018d, park2019}. In our case, an obvious connection is provided by the contemporaneous flares in $\gamma$-ray and radio bands. Due to the timing of its radio variability, we can regard $S1$ as the place of origin of the 2016 $\gamma$-ray outburst. Considering the sampling interval of the VLBA observations (i.e., $\sim$30\,days), the $S1$ flare clearly coincides with the $\gamma$-ray outburst, implying a co-spatial emission region. Such an event can be caused by the passage of a powerful disturbance through a standing feature, which is confirmed visually in Figure~\ref{fig:f7}. We provide the source-frame brightness temperatures for each stationary component in Figure~\ref{fig:f10}. As can be seen, $S1$ reaches $2.3\times10^{12}$\,K, thus implying the dominance of particle energy during the $\gamma$-ray outburst. This strengthens the argument that $S1$ is responsible for the $\gamma$-ray outburst. The plasma in the region of the standing shock is likely to be turbulent \citep{marscher2016}. In this case, particles in $S1$ can be accelerated by second-order Fermi acceleration \citep[e.g.,][]{asano2014} and/or magnetic reconnections \citep[e.g.,][]{sironi2014} in disordered magnetic fields. A moving shock is able to further compress and energize the already excited particles, thus producing strong $\gamma$-ray emission.

Meanwhile, we cannot rule out the possibility of $S2$ being responsible for the 2016 $\gamma$-ray outburst. A polarized knot that enters the $S2$ region is detected just after the $\gamma$-ray outburst. This might be attributed to a shock interaction between $S2$ and a disturbance which would cause a $\gamma$-ray flare in $S2$ (e.g., \citealt{agudo2011a}; see also \citealt{hughes2005}, for discussion of plasma compression resulting in magnetic field enhancement). Such a scenario is natural as the disturbance continues to travel along the jet. However, we noticed several interesting phenomena accompanying this particular disturbance. From MJD 57388 to 57419, the total intensity peak and the polarized knot propagated downstream simultaneously but were located at different positions. Subsequently, the strong radio flare occurred in $S2$ which dominates the estimate for the $\gamma$-ray/radio time delay of $\sim$100 days. Taken together, the observations suggest a large size of the emitting blob \citep[e.g.,][]{lisakov2017}. It has been suggested that the high energy electrons can be confined to a narrow and thin region of a shock front \citep{wehrle2012, marscher2014}. As the moving blob begins to interact with $S2$, a strong $\gamma$-ray emission could have occurred in the shock front which is the injection site \citep[e.g.,][]{agudo2011b}. The strong $S2$ flare reaching up to $\approx$10.5 Jy is remarkable. During the passage of a disturbance in the $S2$ region, there seems to be a huge increase in particle density and/or magnetic field strength that could cause a temporary change of the opacity in the shocked region \citep[e.g.,][]{kravchenko2016}. \citet{lisakov2017} found that one of the flares in the 43\,GHz core (event $B_{7}$ in their nomenclature) peaked when it reached its most downstream position. We find such a pattern for $S2$ in Figure~\ref{fig:f7} (i.e., from MJD 57419 to 57501). This is indicative of that $S2$, like the core, is indeed a recollimation shock.

The association of the $\gamma$-ray emission with the $S1$ or $S2$ regions in the jet ($\geq$~$10^{4}$~$R_{s}$) disfavors the BLR as the source of seed photons. At the distances of $S1$ and $S2$, the BLR is not supposed to be an effective source of seed photons for IC scattering unless the BLR is more extended than expected \citep{leon2013}. Recent studies found indication that the majority of $\gamma$-ray bright FSRQs radiate $\gamma$-rays beyond the BLR region, thus supporting the parsec-scale scenario \citep{costamante2018, meyer2019}. \citet{dotson2012} suggested energy-dependent cooling times for $\gamma$-ray emission produced via infrared (IR) seed photons from the dusty torus. Inverse Compton scattering of IR seed photons occurs in the Thomson regime, whereas the Klein$-$Nishina regime is relevant for higher-energy seed photons (e.g., UV photons from the BLR region), resulting in energy-independent cooling times \citep[see][]{blumenthal1970}. We tried to test this theory by using $\gamma$-ray data binned into very short intervals (e.g., 3 hours). However, insufficient photon statistics prevented the creation of meaningful $\gamma$-ray light curves. More powerful flares or brighter blazars might be suitable for such an analysis. Lower-energy seed photons could originate from the dusty torus or the jet itself (e.g., \citealt{marscher2010, wehrle2016}; see also \citealt{banasinski2018}, for discussion of another emitting blob in the jet that provides seed photons).

\subsection{On the 2017 $\gamma$-ray outburst}
\label{sec:s4.3}

We found a fast and short $\gamma$-ray outburst on MJD 57883 (see Figure~\ref{fig:f2}). The contemporaneous radio flaring activity in both the core and $S1$, however, makes it difficult to interpret the $\gamma$-ray outburst. The total intensity radio maps obtained just after the $\gamma$-ray outburst suggest a strong disturbance localized between the core and $S1$. As discussed before, we assume that the core was in the onset of its flare while $S1$ was already in the middle of its flare at the time of the $\gamma$-ray outburst. This can be explained by the presence of multiple moving disturbances. The polarization image of the jet observed in MJD 57886 shows a weak polarized knot near the core which indicates the emergence of a new disturbance. Hence, we consider two propagating disturbances around the time of the $\gamma$-ray outburst: one interacting with the core ($K1$) and one interacting with $S1$ ($K2$). This picture is consistent with the jet features observed on MJD 57971: (1) for $K1$, the total intensity peak located at $S1$, and (2) for $K2$, the enhanced flux of $S2$. An increasing trend in the core size from MJD 57886 to 57937 might further support the presence of $K1$. 

Figure~\ref{fig:f2} shows that the 2017 $\gamma$-ray outburst lasted for a short time (probably less than a day), very different from the 2016 outburst. The short time scale suggests a small emitting region \citep[e.g.,][]{petropoulou2015}, implying a region in the jet located closer to the central engine \citep{tavecchio2010}. The relatively low apparent component speeds around the $\gamma$-ray outburst further support this idea \citep[][]{rani2018}. We also note that the $\gamma$-ray outburst triggered a mm-wavelength flare in the core that might have been accompanied by the emergence of a polarized knot nearby. Hence, the core is likely to be responsible for the $\gamma$-ray outburst. Indeed, such a timing of events -- a $\gamma$-ray outbursts occurring at the very beginning of a radio flare -- is common in blazars \citep[e.g.,][]{marscher2016, lisakov2017}. The $\gamma$-ray outburst could have been caused by the interaction of the standing shock and a strong disturbance (i.e., $K1$) propagating down the jet, in the same manner as explained in Section~\ref{sec:s4.2}. In addition, the brightness temperature of the core is peaking at the time of the $\gamma$-ray outburst, which further supports this scenario.

\begin{figure}
 \includegraphics[width=\columnwidth]{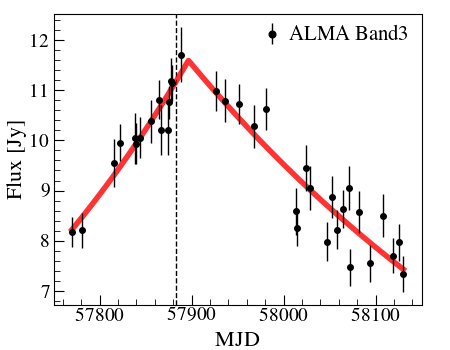}
 \centering 
 \caption{The 3\,mm ALMA light curve around the time of the 2017 $\gamma$-ray outburst (vertical dashed line). A model combining an exponential rise with a subsequent exponential decay (as introduced by \citealt{valtaoja1999}, but without fixing the decay timescale) is indicated by the red curve.
}
 \label{fig:f11}
\end{figure}

It is worth noting that there seems to be a discrepancy in position between the polarized knot and the core component. This might be due to variable opacity in the core region \citep[e.g.,][]{lisakov2017}, when the core region becomes partially optically thin temporarily. Indeed, the 3-mm-to-7-mm spectral index varied rapidly around the time of the $\gamma$-ray outburst (see Figure~\ref{fig:f1}). Modelling the minor ALMA flare associated with the $\gamma$-ray outburst as a combination of an exponentially rising and an exponentially declining part returns a 3\,mm peak at MJD 57897$\pm$7 (see Figure~\ref{fig:f11}). This corresponds to a time delay of about 40\,days relative to the 7\,mm core flare observed on MJD 57937. The non-zero time lag implies optically thick emission, though the source of the ALMA flare is unknown (i.e., among the core/$S1$/$S2$). Nevertheless, this is relevant with respect to the result of \citet{chidiac2016} who found a time delay of $\sim$50 days between the 43 and 230\,GHz light curves of 3C\,273. In short, we are not able to determine the opacity condition in the core during the $\gamma$-ray outburst.

In the conventional view, the core shift might play a role, though it is expected to occur at very small sub-milliarcsec scales at wavelengths below 7\,mm \citep[][]{volvach2013, lisakov2017}. Interestingly, the 2017 $\gamma$-ray outburst coincides with not only the onset of the 7\,mm flare in the core, but also with the peak of the 3\,mm flare at the same time. Thus, we suggest that a moving blob (i.e., $K1$) started interaction with the 7\,mm core, while the blob (probably its tail part) still passed through the 3\,mm core. This scenario could explain the polarized feature that appeared on MJD 57886 upstream from the 7\,mm core. Furthermore, a passage of $K1$ through the 3\,mm core can explain the contemporaneous $\gamma$-ray/3\,mm flares \citep[e.g.,][]{jorstad2013, wehrle2016}. Given our observations of the $\gamma$-ray outburst, however, we cannot definitively locate the origin of the $\gamma$-ray outburst between the 3 and 7\,mm core. Considering the mm-wavelength core as the $\gamma$-ray production site, IR emission from either the dusty torus \citep[e.g.,][]{marscher2014}, stationary knots \citep[e.g.,][]{wehrle2016}, or jet sheath are candidate sources of seed photons, as discussed in the previous section.

\subsection{$\gamma$-ray spectra}
\label{sec:s4.4}

Overall, spectral index values around $-$2.8 (Figure~\ref{fig:f3}) are close to the value reported in \citet{harris2012}. Given the typical photon indices of $\gamma$-ray bright FSRQs, which is around $-$2.4, however, 3C\,273 shows a softer photon index \citep[see][]{harris2012, linford2012}. Although 3C\,273 was less $\gamma$-ray bright than those extreme FSRQs (e.g., 3C~279 and 3C~454.3), it seems that the difference in photon index values is a sign of different source physics, like a spectral break or additional emitting components (\citealt{harris2012, hess2018}; see also \citealt{costamante2018}, for the exception of 3C\,273 at constraining high/low states). We also note that our parsec-scale scenario supports what \citet{harris2012} reported: $\gamma$-ray absorption by BLR photons is disfavored in $\gamma$-ray bright FSRQs.

$\gamma$-ray outbursts, accompanied by variability of the $\gamma$-ray spectral index, have been reported in many previous studies \citep{rani2013b, rani2018, kim2018d, ding2019}. The usual combination is an increase of $\gamma$-ray flux density and  hardening of photon indices. Unfortunately, we are not able to draw a clear quantitative conclusion from our data due to insufficient photon statistics. However, a simple visual inspection of the monthly binned photon indices in Figure~\ref{fig:f4} suggests such an evolution for both the 2016 and 2017 $\gamma$-ray outbursts. The peak of the 2016 $\gamma$-ray outburst lags about $\sim$40 days behind a local maximum in the spectral index time series on MJD 57338, whereas the $\gamma$-ray peak of the 2017 outburst coincides with a local photon index peak on MJD 57878. If we identify each of the local photon index peaks with the first shock interaction between a strong disturbance propagating down the jet and the mm-wave core, we can make a smooth connection to our parsec-scale scenario for both $\gamma$-ray outbursts.

The overall spectral behavior of the monthly binned $\gamma$-rays shown in Figure~\ref{fig:f3} indicates a transition from a softer-when-brighter state to a harder-when-brighter state at a critical $\gamma$-ray flux of around $1.03\times10^{-7}$\,ph\,cm$^{-2}$\,s$^{-1}$, though the softer-when-brighter trend is only marginally significant \citep[see also][for similar trends in other sources]{abdo2010b, paliya2015, kim2018d}. This transition can be explained by a balance between acceleration and cooling of relativistic particles, which in turn implies cooling is dominant for the softer-when-brighter trend. Such a transition might also be related to a shift of the IC peak in the SED \citep{shah2019}. As the source power increases during strong flares, the IC peak can be shifted to higher energies, therefore resulting in a harder-when-brighter trend.

\section{Conclusions}
\label{sec:s5}

Detailed analysis of the $\gamma$-ray and radio activity of the blazar 3C\,273 during 2015--2019 enables us to address the nature of flaring activity and its connection to the parsec scale jet. Throughout our observations, 3C\,273 experienced two significant $\gamma$-ray outbursts, accompanied by strong flaring activity at mm-wavelengths. We identified three quasi-stationary components in the compact inner region of the jet: the core, $S1$, and $S2$. These features are confirmed to be the main sources of the observed mm-wavelength emission, exhibiting powerful flaring variability. Joint analysis of the $\gamma$-ray and radio data reveals strong correlation between the two energy regimes, and provides strong evidence for the inner jet region to be the production site of $\gamma$-ray emission. Overall, the observed behavior can be explained by a scenario in which the radio and $\gamma$-ray flares are produced in the core--$S1$--$S2$ region by moving disturbances and their interaction with the standing shocks. The emergence of notable polarized knots and high brightness temperatures (up to $\sim\,10^{12}\,K$) around the times of the $\gamma$-ray outbursts further support such a scenario. As the blob propagates down the jet, it causes changes in the physical conditions in the ambient flow, resulting in the observed spatial displacements of stationary components (the core and $S2$) and variations in spectral properties.

\begin{acknowledgements}
This work makes use of public Fermi data obtained from \textit{Fermi} Science Support Center (FSSC). This paper makes use of the following ALMA data: ADS/JAO.ALMA\#2011.0.00001.CAL. ALMA is a partnership of ESO (representing its member states), NSF (USA) and NINS (Japan), together with NRC (Canada), MOST and ASIAA (Taiwan), and KASI (Republic of Korea), in cooperation with the Republic of Chile. The Joint ALMA Observatory is operated by ESO, AUI/NRAO and NAOJ. This study makes use of 43 GHz VLBA data from the VLBA-BU Blazar Monitoring Program (VLBA-BU-BLAZAR; http://www.bu.edu/blazars/VLBAproject.html), funded by NASA through the Fermi Guest Investigator Program. The VLBA is an instrument of the National Radio Astronomy Observatory. The National Radio Astronomy Observatory is a facility of the National Science Foundation operated by Associated Universities, Inc. Dae-Won Kim acknowledges support from the National Research Foundation of Korea (NRF) through fellowship 2019R1A6A3A13095962. Sascha Trippe and Dae-Won Kim acknowledges support from the NRF grant 2019R1F1A1059721. EVK acknowledges support through the contract ASI-INAF 2015-023-R.1-2019.

\end{acknowledgements}

\begin{appendix} 
\section{Appendix}

\longtab[1]{
\begin{longtable}{cccccccccccc}
\caption{Parameters of the 43\,GHz components obtained from Gaussian model fits.} \\
\hline
\hline
Date & MJD & Comp.$^a$ & Flux & Distance$^b$ & Angle$^c$ & Size & B$_\mathrm{maj}^d$ & B$_\mathrm{min}^d$ & B$_\mathrm{PA}^d$ & rms$^e$ & Peak$^f$ \\
 & & & (Jy) & (mas) & ($^{\circ}$) & (mas) & (mas) & (mas) & ($^{\circ}$) & (mJy/beam) & (Jy/beam) \\
\hline
\endfirsthead

\caption{Continued.} \\
\hline
\hline
Date & MJD & Comp.$^a$ & Flux & Distance$^b$ & Angle$^c$ & Size & B$_\mathrm{maj}^d$ & B$_\mathrm{min}^d$ & B$_\mathrm{PA}^d$ & rms$^e$ & Peak$^f$ \\
 & & & (Jy) & (mas) & ($^{\circ}$) & (mas) & (mas) & (mas) & ($^{\circ}$) & (mJy/beam) & (Jy/beam) \\
\hline
\endhead


2015-12-05 & 57361 & Core & 2.24 & 0.00 & $-$ & 0.09 & 0.45 & 0.15 & $\mathrm{-}$12.70 & 4.80 & 4.16 \\
 & & S1 & 3.76 & 0.16 & $\mathrm{-}$145.03 & 0.05 &  &  &  &  & \\
  & & S2 & 1.19 & 0.36 & $\mathrm{-}$135.78 & 0.16 &  &  &  &  & \\
   & & J1 & 0.72 & 0.71 & $\mathrm{-}$130.74 & 0.13 &  &  &  &  & \\
    & & J2 & 0.67 & 0.87 & $\mathrm{-}$139.77 & 0.12 &  &  &  &  & \\
     & & J3 & 0.70 & 0.91 & $\mathrm{-}$131.31 & 0.08 &  &  &  &  & \\
      & & J4 & 0.05 & 1.60 & $\mathrm{-}$131.37 & 0.09 &  &  &  &  & \\
       & & J5 & 0.21 & 3.02 & $\mathrm{-}$139.39 & 0.36 &  &  &  &  & \\
        & & J6 & 0.12 & 3.12 & $\mathrm{-}$133.67 & 0.12 &  &  &  &  & \\
         & & J7 & 0.29 & 5.93 & $\mathrm{-}$140.29 & 0.92 &  &  &  &  & \\
2016-01-01 & 57388 & Core & 2.02 & 0.00 & $-$ & 0.08 & 0.43 & 0.17 & $\mathrm{-}$7.84 & 10.76 & 5.31 \\
 & & S1 & 4.77 & 0.16 & $\mathrm{-}$145.78 & 0.04 &  &  &  &  & \\
  & & S2 & 1.09 & 0.37 & $\mathrm{-}$135.58 & 0.15 &  &  &  &  & \\
   & & J1 & 1.76 & 0.87 & $\mathrm{-}$134.01 & 0.22 &  &  &  &  & \\
    & & J2 & 0.08 & 1.31 & $\mathrm{-}$139.52 & 0.13 &  &  &  &  & \\
     & & J3 & 0.29 & 3.10 & $\mathrm{-}$136.72 & 0.48 &  &  &  &  & \\
      & & J4 & 0.22 & 5.95 & $\mathrm{-}$140.31 & 0.82 &  &  &  &  & \\
2016-02-01 & 57419 & Core & 1.74 & 0.00 & $-$ & 0.04 & 0.39 & 0.14 & $\mathrm{-}$12.80 & 17.55 & 6.73 \\
 & & S1 & 2.44 & 0.15 & $\mathrm{-}$130.03 & 0.02 &  &  &  &  & \\
  & & S2 & 6.23 & 0.27 & $\mathrm{-}$141.26 & 0.06 &  &  &  &  & \\
   & & J1 & 1.21 & 0.46 & $\mathrm{-}$133.62 & 0.19 &  &  &  &  & \\
    & & J2 & 1.56 & 0.95 & $\mathrm{-}$134.40 & 0.15 &  &  &  &  & \\
     & & J3 & 0.60 & 1.08 & $\mathrm{-}$130.51 & 0.07 &  &  &  &  & \\
      & & J4 & 0.08 & 1.34 & $\mathrm{-}$140.46 & 0.11 &  &  &  &  & \\
       & & J5 & 0.47 & 3.24 & $\mathrm{-}$135.71 & 0.59 &  &  &  &  & \\
2016-03-19 & 57466 & Core & 1.82 & 0.00 & $-$ & 0.07 & 0.42 & 0.17 & $\mathrm{-}$6.95 & 7.95 & 8.39 \\
 & & S1 & 1.61 & 0.15 & $\mathrm{-}$132.52 & 0.07 &  &  &  &  & \\
  & & S2 & 8.50 & 0.32 & $\mathrm{-}$140.96 & 0.07 &  &  &  &  & \\
   & & J1 & 0.38 & 0.44 & $\mathrm{-}$127.67 & 0.10 &  &  &  &  & \\
    & & J2 & 0.19 & 0.72 & $\mathrm{-}$133.32 & 0.06 &  &  &  &  & \\
     & & J3 & 0.43 & 0.90 & $\mathrm{-}$128.88 & 0.18 &  &  &  &  & \\
      & & J4 & 1.99 & 1.12 & $\mathrm{-}$133.70 & 0.20 &  &  &  &  & \\
       & & J5 & 0.26 & 3.39 & $\mathrm{-}$136.05 & 0.40 &  &  &  &  & \\
        & & J6 & 0.26 & 6.18 & $\mathrm{-}$140.76 & 0.84 &  &  &  &  & \\
2016-04-23 & 57501 & Core & 1.86 & 0.00 & $-$ & 0.03 & 0.37 & 0.15 & $\mathrm{-}$3.62 & 17.22 & 9.10 \\
 & & S1 & 1.54 & 0.15 & $\mathrm{-}$132.81 & 0.08 &  &  &  &  & \\
  & & S2 & 10.49 & 0.37 & $\mathrm{-}$139.47 & 0.09 &  &  &  &  & \\
   & & J1 & 0.61 & 0.62 & $\mathrm{-}$129.17 & 0.29 &  &  &  &  & \\
    & & J2 & 0.61 & 0.99 & $\mathrm{-}$131.57 & 0.14 &  &  &  &  & \\
     & & J3 & 2.06 & 1.22 & $\mathrm{-}$132.81 & 0.20 &  &  &  &  & \\
      & & J4 & 0.18 & 3.49 & $\mathrm{-}$136.15 & 0.45 &  &  &  &  & \\
 \hline
2017-04-16 & 57859 & Core & 2.05 & 0.00 & $-$ & 0.07 & 0.40 & 0.15 & $\mathrm{-}$8.33 & 7.04 & 1.93 \\
 & & S1 & 1.11 & 0.20 & $\mathrm{-}$135.43 & 0.09 &  &  &  &  & \\
  & & S2 & 0.25 & 0.40 & $\mathrm{-}$141.68 & 0.12 &  &  &  &  & \\
   & & J1 & 1.07 & 0.81 & $\mathrm{-}$139.24 & 0.23 &  &  &  &  & \\
    & & J2 & 0.03 & 1.06 & $\mathrm{-}$141.93 & 0.12 &  &  &  &  & \\
     & & J3 & 0.72 & 1.28 & $\mathrm{-}$130.05 & 0.26 &  &  &  &  & \\
      & & J4 & 0.26 & 1.75 & $\mathrm{-}$137.87 & 0.18 &  &  &  &  & \\
       & & J5 & 0.52 & 2.00 & $\mathrm{-}$132.47 & 0.25 &  &  &  &  & \\
2017-05-13 & 57886 & Core & 2.32 & 0.00 & $-$ & 0.04 & 0.35 & 0.15 & $\mathrm{-}$4.92 & 15.88 & 3.34 \\
 & & S1 & 2.55 & 0.14 & $\mathrm{-}$144.45 & 0.06 &  &  &  &  & \\
  & & S2 & 0.99 & 0.34 & $\mathrm{-}$138.72 & 0.12 &  &  &  &  & \\
   & & J1 & 1.09 & 0.77 & $\mathrm{-}$139.49 & 0.27 &  &  &  &  & \\
    & & J2 & 0.83 & 1.03 & $\mathrm{-}$140.71 & 0.25 &  &  &  &  & \\
     & & J3 & 0.62 & 1.27 & $\mathrm{-}$129.14 & 0.16 &  &  &  &  & \\
      & & J4 & 0.88 & 1.78 & $\mathrm{-}$134.10 & 0.37 &  &  &  &  & \\
       & & J5 & 0.48 & 2.19 & $\mathrm{-}$133.86 & 0.18 &  &  &  &  & \\
2017-06-08 & 57912 & Core & 3.72 & 0.00 & $-$ & 0.07 & 0.38 & 0.16 & $\mathrm{-}$2.67 & 13.40 & 5.04 \\
 & & S1 & 3.60 & 0.15 & $\mathrm{-}$142.58 & 0.05 &  &  &  &  & \\
  & & S2 & 1.26 & 0.32 & $\mathrm{-}$137.83 & 0.13 &  &  &  &  & \\
   & & J1 & 0.63 & 0.78 & $\mathrm{-}$142.68 & 0.14 &  &  &  &  & \\
    & & J2 & 1.01 & 0.97 & $\mathrm{-}$138.03 & 0.21 &  &  &  &  & \\
     & & J3 & 0.67 & 1.26 & $\mathrm{-}$129.38 & 0.23 &  &  &  &  & \\
      & & J4 & 0.66 & 1.54 & $\mathrm{-}$131.76 & 0.20 &  &  &  &  & \\
       & & J5 & 0.32 & 1.92 & $\mathrm{-}$139.27 & 0.11 &  &  &  &  & \\
        & & J6 & 0.76 & 2.12 & $\mathrm{-}$133.16 & 0.25 &  &  &  &  & \\
2017-07-03 & 57937 & Core & 4.41 & 0.00 & $-$ & 0.10 & 0.44 & 0.15 & $\mathrm{-}$12.60 & 11.50 & 4.87 \\
 & & S1 & 3.67 & 0.17 & $\mathrm{-}$145.42 & 0.05 &  &  &  &  & \\
  & & S2 & 0.93 & 0.32 & $\mathrm{-}$140.05 & 0.12 &  &  &  &  & \\
   & & J1 & 0.08 & 0.59 & $\mathrm{-}$147.08 & 0.09 &  &  &  &  & \\
    & & J2 & 0.71 & 0.83 & $\mathrm{-}$141.58 & 0.15 &  &  &  &  & \\
     & & J3 & 0.53 & 1.02 & $\mathrm{-}$138.43 & 0.14 &  &  &  &  & \\
      & & J4 & 0.54 & 1.20 & $\mathrm{-}$129.23 & 0.25 &  &  &  &  & \\
       & & J5 & 0.76 & 1.55 & $\mathrm{-}$132.29 & 0.28 &  &  &  &  & \\
        & & J6 & 0.36 & 1.95 & $\mathrm{-}$138.81 & 0.24 &  &  &  &  & \\
         & & J7 & 0.56 & 2.18 & $\mathrm{-}$133.08 & 0.23 &  &  &  &  & \\
2017-08-06 & 57971 & Core & 1.59 & 0.00 & $-$ & 0.04 & 0.36 & 0.13 & $\mathrm{-}$10.5 & 9.36 & 2.91 \\
 & & S1 & 2.37 & 0.12 & $\mathrm{-}$127.25 & 0.06 &  &  &  &  & \\
  & & S2 & 2.00 & 0.26 & $\mathrm{-}$139.42 & 0.04 &  &  &  &  & \\
   & & J1 & 0.26 & 0.41 & $\mathrm{-}$137.03 & 0.05 &  &  &  &  & \\
    & & J2 & 0.56 & 0.90 & $\mathrm{-}$140.27 & 0.19 &  &  &  &  & \\
     & & J3 & 0.61 & 1.19 & $\mathrm{-}$134.08 & 0.30 &  &  &  &  & \\
      & & J4 & 0.29 & 1.47 & $\mathrm{-}$128.01 & 0.24 &  &  &  &  & \\
       & & J5 & 0.52 & 1.87 & $\mathrm{-}$135.26 & 0.34 &  &  &  &  & \\
        & & J6 & 0.33 & 2.31 & $\mathrm{-}$133.10 & 0.29 &  &  &  &  & \\

\hline
\multicolumn{12}{l}{$^a$ Higher Jx numbers correspond to larger downstream distances from the core.}\\
\multicolumn{12}{l}{$^b$ Distance from the core.}\\
\multicolumn{12}{l}{$^c$ Position angle relative to core component.}\\
\multicolumn{12}{l}{$^d$ FWHM of the elliptical beam: the major and minor axis, and the inclination angle of the major axis with respect to the North.}\\
\multicolumn{12}{l}{$^e$ rms noise of residual map.}\\
\multicolumn{12}{l}{$^f$ Map peak of CLEAN map.}\\
\label{app:a1}
\end{longtable}
}

\end{appendix}

\end{document}